\documentclass[letterpaper, 11pt]{article}
\usepackage{amsmath}
\usepackage{amssymb}
\usepackage{graphicx}
\usepackage{subcaption}
\usepackage{bm}
\usepackage{cancel}
\usepackage{verbatim} 
\usepackage[margin=0.5in]{geometry}
\usepackage[title]{appendix}
\usepackage{hhline}
\usepackage{bibentry}
\usepackage{setspace}
\usepackage{algorithm}
\usepackage{algpseudocodex}
\usepackage{comment}
\usepackage{mathtools}

\begin{document}

\title{Superstep wavefield propagation}
\author{Tamas Nemeth$^1$, Kurt Nihei$^1$, Alex Loddoch$^1$, Anusha Sekar$^1$, Ken Bube$^2$, 
\\ John Washbourne$^1$, Luke Decker$^1$, Sam Kaplan$^1$,
Chunling Wu$^1$, \\Andrey Shabelansky$^1$, Milad Bader$^1$, Ovidiu Cristea$^1$ and Ziyi Yin$^3$}
\date{$^1$ Chevron Technical Center, a division of Chevron U.S.A. Inc.\\
      $^2$ University of Washington\\
      $^3$ Georgia Institute of Technology\\
      May 31, 2024}



\maketitle

\begin{abstract}
This paper describes how to propagate wavefields for arbitrary numbers of traditional time steps in a single step, called a superstep. 
We show how to construct operators that accomplish this task for finite-difference time domain schemes, including 
temporal first-order schemes in isotropic, anisotropic and elastic media, as well as temporal second-order schemes for acoustic media.
This task is achieved by implementing a computational tradeoff differing from traditional single step wavefield propagators 
by precomputing propagator matrices for each model location for $k$ timesteps (a superstep) and using these propagator
matrices to advance the wavefield $k$ time steps at once. 
This tradeoff separates the physics of the propagator matrix computation from the computer science of wavefield propagation 
and allows each discipline to provide their optimal modular solutions.
\end{abstract}


\section{Introduction}
Traditional wavefield propagation using finite-difference time domain (FD) schemes advances the wavefield one timestep at a time 
by performing a matrix-vector multiplication of the current wavefields with the combination of 
material properties (such as velocity or density) in a pattern described by the FD kernel. 
The accuracy, stability and dispersion of such FD schemes have been studied extensively 
(Hairer et al. \cite{hairer}, Chapman \cite{chapman}, Durran \cite{durran}, Bube et al. \cite{bn}, 
Bube et al. \cite{bn2}, Bube et al. \cite{bw}, Washbourne et al. \cite{bw2}) 
and are well understood. 
Today, most modeling and imaging applications are built on linear or linearized FD schemes from 
acoustic forward modeling to vertical and tilted transverse isotropy (VTI, TTI) and elastic modeling. 

Traditional recursive time steps with such a FD scheme operationally correspond to a repeated matrix-vector 
multiplication, although the matrices and vectors may not be explicitly formed. 
Such schemes can be viewed as a series of matrices (one for each time step) applied to a vector representing the current wavefield state. 
It is widely accepted, especially for large matrices, that computationally it is better to carry out this computation sequentially 
from “right” to “left”; start by multiplying the initial wavefield vector on the right 
with the matrix for the first time step resulting in a new vector; applying the next timestep matrix 
to the new vector, until all matrix-vector multiplications are executed.

Performing operations in this manner reduces computational cost, as 
these traditional schemes map well onto current hardware (CPU, GPU) where the wavefield propagation problem is 
considered as a mostly compute-bound computational problem. 
This established process performs well when most of these jobs are run on high-end hardware nodes and 
storage systems in the data centers owned by the organizations themselves. 
With the availability of broad classes of compute nodes and storage options in the cloud, hardware choices 
have become more diverse and, correspondingly, the associated cost structure has become more complex but visible. 
For wavefield propagation using the cloud infrastructure, the overall runtime nowadays mostly depends on the sophistication of managing 
the storage and IO of the wavefields and earth model material parameters both in the memory of compute nodes and online storage,
and on the agility of the scheduling systems for virtual machine compute nodes.
These developments necessitate the re-examination of the traditional computational paradigm for 
wavefield propagation.

This paper presents numerical formulations for wavefield propagation where the tradeoff for compute and storage 
is different from the traditional formulation and the algorithmic components are more general. 
These formulations enable us to recast the process of wavefield propagation as two distinct steps 
of “Precompute” and “Compute”. 
During Precompute, the common reusable components, the propagator matrices (PM) are calculated and stored. 
This Precompute stage is separate from the subsequent Compute step and can therefore be executed on separate systems 
possibly with a different hardware architecture.
During the actual Compute stage, the precomputed PMs are used and re-used 
to achieve wavefield propagation.

There were several earlier efforts to use Precompute for part of the wavefield propagation. 
Notable is the checkpointing scheme of Symes \cite{symes} where wavefields are systematically saved 
to match the forward and adjoint wavefields. 
Kole \cite{kole} used a matrix exponential approach to derive unconditionally stable algorithms
to approximate time evolution (numerical integration of the matrix exponential power series) by 
operator splitting and one-step Chebyshev polynomials. 
Araujo and Pestana \cite{pestana} also used matrix exponential approaches to model the
time evolution of the first-order linear acoustic and elastic wave equations.
In a different domain, stencil computations were approximated with Gaussian fitting and 
then these Gaussians were used for further propagation (Ahmad et al. \cite{ahma}). 
Jun \cite{jun} and Vamaraju et al. \cite{janaki} formulated seismic imaging as a repeated 
application of system matrices.
Vdovina et al. \cite{vdovina}, Korostyshevskaya et al. \cite{korosh} and Gibson et al. \cite{gibson}
developed an operator upscaling method to capture the finer subgrid effects 
on a coarser domain without updating the wavefields on a finer scale.

We have found that the optimal Precompute is to calculate the superstepping propagator matrices associated with 
each gridpoint in the spatial discretization of the wave propagation domain
as these objects are the most re-usable and can be applied to wavefields at any timesteps. 
The corresponding Compute is the integration of the PMs with the wavefields at each grid point 
and the output is the updated wavefield. 

The key observation to superstepping is to re-arrange the FD schemes in such a way that current and 
previous wavefields are related to future and subsequent wavefields via a square system matrix. 
This way the square matrix can be raised to the $k$th power corresponding to superstepping of $k$ traditional time steps. 
When the initial wavefield is a delta function for each spatial location, we can precompute the Green's functions 
at each location to the $k$th traditional time steps. 
These precomputed PMs become impulse response filters at these locations. 
Then we apply these filters to the actual wavefields to advance the wavefield by $k$ steps (a superstep). 

Similar ideas have appeared in different contexts. 
Seismic interferometry uses Green's functions to reconstruct wavefields from other wavefields 
(Wapenaar et al. \cite{Wapenaar2006}, Schuster \cite{schuster}, Wapenaar et al. \cite{Wapenaar2016}). 
Superstepping can be considered as a special case of seismic interferometry. 
Similarly, Marchenko methods (Slob et al. \cite{slob}, Lomas and Curtis \cite{lomas}, Wapenaar \cite{Wapenaar2022}) 
reconstruct subsurface Green's functions and use them for imaging.
Wapenaar \cite{Wapenaar2022} especially, shows the interrelation and differences between Green's functions, propagator matrices and Marchenko focusing functions.
Since a propagator matrix depends only on the medium parameters between the two time levels where defined in the medium,
we will use the terminology of propagator matrices thereon in accordance with Wapenaar \cite{Wapenaar2022}.

Recently, many novel machine learning techniques were developed that train to predict system states 
continuously from current time to up to a given future time. 
These methods, including Fourier Neural Operators \cite{fno}, \cite{fno_felix} use the inherent nonlinearity 
in the neural network to provide a model for inference. 
Although both machine learning methods and superstepping can predict future states of the system, 
superstepping is based on the linear properties of the system and is not an approximation to the traditional time stepping schemes.
Some neural network methods provide models to some tasks used in traditional wavefield propagation 
such as boundary condition prediction (Kaur et al. \cite{kaur}). 
In this work we show that these tasks can also be precomputed as they are part of the linear system matrix.

As seen in later sections, the Precompute stage can be identified in incorporating physics into the Green's functions 
and the Compute stage can be viewed as a computer science effort to match computational tasks 
to the given computing resources. 
For the heterogeneous cloud-based compute and storage environment, this match may be best accomplished 
via efficient domain-specific languages (DSL). 
Earlier, Charlesworth et al. \cite{fddsl} developed a DSL for finite-difference modeling across CPU, FPGA and GPU hardware. 
Recently the Devito framework \cite{devito} is showing promise as a DSL for FD stencil computations. 
We hope that the superstepping scheme will provide yet another DSL abstraction with the benefit 
that the Compute stage involves only computer science considerations.

The manuscript is organized as follows.
In the Theory section we give a concise description of superstepping for both first- and second-order systems
of wave propagation and show that simply an outcome of precomputing the $k$th power of the square is the 
propagator matrix.
In the System matrix representations as propagator matrices section we show that the propagator matrices
are composed of PMs for each row of the matrix and this greatly reduces the non-zero elements
of these matrices.
In the Superstepping computational model section we describe the numerical aspects of both Precompute and Compute stages.
In the Synthetic examples section we demonstrate superstepping on a Marmousi 2D example and show
the treatment of boundary reflections in the Green's functions.
The Discussion section emphasizes one of the main results, that the Compute stage is 
is the stage for computational and algorithmic implementations rather than physics.
Finally, in the Appendices section we give detailed examples of formulating the first and second-order 
acoustic, VTI anisotropic and elastic wave equations in terms of superstepping.

\section{Theory}
Many physical phenomena can be expressed via partial differential equations (PDE) where state variables change over time and space.
An important class of these partial differential equations is shown in the next set of equations 
where state variables $\mathbf{u}$ and $\mathbf{v}$ are related to each other's temporal changes
and typically indicate the velocity and stress relations in the medium. 
The state variables $\mathbf{u}$ and $\mathbf{v}$ in the physical domain can be both multi-dimensional and multi-component quantities.

\subsection{First-order partial differential equations}
A typical first order system of equations with 2 state variables for wave propagation can be written in a matrix equation form as
\begin{eqnarray} \label{eq:fun1}
  \left[
  \begin{array} {c}
  \mathbf{\dot{u}} \\
  \mathbf{\dot{v}} 
  \end{array}
  \right]
  =
  \begin{bmatrix}
    0 & A \\
    B & 0 
  \end{bmatrix}
  \left[
  \begin{array} {c}
  \mathbf{u} \\
  \mathbf{v}
  \end{array}
  \right] ~,
\end{eqnarray}
where $\mathbf{\dot{u}}$ and $\mathbf{\dot{v}}$ are the temporal derivatives of state variables 
$\mathbf{u}$ and $\mathbf{v}$, typically wavefields of particle velocities and stress in our case. 
Matrices $A$ and $B$ incorporate material coefficients and partial derivative operators such as $\nabla$ and $\nabla \cdot$ operators. 
Note that $A$ and $B$ are off the main diagonal of the matrix indicating that the spatial variations of 
one quantity affects the temporal variations of the other one. 
This is typical for many PDEs in wave propagation and we will provide specific examples later. 
There are more complex systems of first order equations for other physical formulations and 
we will discuss later as well.

Equation \ref{eq:fun1} can be discretized using the symplectic Euler method:
\begin{eqnarray} \label{eq:fun2}
  \mathbf{u}_{n+1} & = & \mathbf{u}_n + A \, \mathbf{v}_n ~, \nonumber \\
  \mathbf{v}_{n+1} & = & \mathbf{v}_n + B \, \mathbf{u}_{n+1} ~,
\end{eqnarray}
where $\mathbf{\dot{u}} = (\mathbf{u}_{n+1} - \mathbf{u}_n) / \Delta t$, 
$\mathbf{\dot{v}} = (\mathbf{v}_{n+1} - \mathbf{v}_n) / \Delta t$, and $\Delta t$ is absorbed into $A$ and $B$ 
for simplicity. The index $n$ denotes the discretized time step. 
Sometimes the $\mathbf{u}$ variable is staggered compared to the $\mathbf{v}$ variable and
$\mathbf{\dot{u}}$ is instead equal to $(\mathbf{u}_{n+1/2} - \mathbf{u}_{n-1/2}) / \Delta t$. 
The staggered nature of time steps for the variables is implied when needed. 
Equation \ref{eq:fun2} can be expressed in a matrix form as
\begin{eqnarray} \label{eq:fun3}
  \left[
  \begin{array} {c}
  \mathbf{u}_{n+1} \\
  \mathbf{v}_{n+1} 
  \end{array}
  \right]
  =
  \begin{bmatrix}
    I & 0 \\
    B & I
  \end{bmatrix}
  \begin{bmatrix}
  I & A \\
  0 & I
  \end{bmatrix}
  \left[
  \begin{array} {c}
  \mathbf{u}_n \\
  \mathbf{v}_n 
  \end{array}
  \right] 
  =
  \begin{bmatrix}
  I & A \\
  B & I+BA
  \end{bmatrix}
  \left[
  \begin{array} {c}
  \mathbf{u}_n \\
  \mathbf{v}_n 
  \end{array}
  \right] ~,
\end{eqnarray}
where the Lie-Trotter split-step update matrices
  $\begin{bmatrix}
  I & 0 \\
  B & I
  \end{bmatrix}$ and  
  $\begin{bmatrix}
  I & A \\   \nonumber
  0 & I      \nonumber
  \end{bmatrix} $
describe the update from $\mathbf{u}_n$ to $\mathbf{u}_{n+1}$ and from $\mathbf{v}_n$ to $\mathbf{v}_{n+1}$. 
These split-step matrices are invertible
  $\begin{bmatrix}
  I & 0 \\
  B & I
  \end{bmatrix}^{-1}
  =
  \begin{bmatrix}
  I & 0 \\
  -B & I
  \end{bmatrix} $ , 
  $\begin{bmatrix}
  I & A \\
  0 & I
  \end{bmatrix}^{-1}
  =
  \begin{bmatrix}
  I & -A \\
  0 & I
  \end{bmatrix}$,
and the inverse split-step update matrices describe the update from $\mathbf{u}_n$ to $\mathbf{u}_{n-1}$ and 
from $\mathbf{v}_n$ to $\mathbf{v}_{n-1}$ indicating forward and backward time symmetry.

The update is a 2-step process for this system with 2 classes of variables.
The progression of the update in terms of the resulting vectors can be shown as
\begin{eqnarray} \label{eq:fun8}
  \left[
  \begin{array} {c}
  \mathbf{u}_{n} \\
  \mathbf{v}_{n} 
  \end{array}
  \right] \Rightarrow
  \left[
  \begin{array} {c}
  \mathbf{u}_{n+1} \\
  \mathbf{v}_{n}
  \end{array}
  \right] \Rightarrow
  \left[
  \begin{array} {c}
  \mathbf{u}_{n+1} \\
  \mathbf{v}_{n+1} 
  \end{array}
  \right] \Rightarrow
  \left[
  \begin{array} {c}
  \mathbf{u}_{n+2} \\
  \mathbf{v}_{n+1} 
  \end{array}
  \right] \Rightarrow
  \left[
  \begin{array} {c}
  \mathbf{u}_{n+2} \\
  \mathbf{v}_{n+2} 
  \end{array}
  \right] 
\end{eqnarray} 
From equation \ref{eq:fun8} it is obvious that repeated applications of these matrices constitute time propagation of these state variables.
This can be expressed by raising the propagator matrices to the $k$th power for $k$ time step advancement, as shown in
\begin{eqnarray} \label{eq:fun9}
  \left[
  \begin{array} {c}
  \mathbf{u}_{n+k} \\
  \mathbf{v}_{n+k} 
  \end{array}
  \right] 
  =
  \begin{bmatrix}
  I & A \\
  B & I+BA
  \end{bmatrix}^k
  \left[
  \begin{array} {c}
  \mathbf{u}_n \\
  \mathbf{v}_n 
  \end{array}
  \right] ~.
\end{eqnarray}
Equation \ref{eq:fun9} implies that we can calculate the $\mathbf{u}_{n+k}, \mathbf{v}_{n+k}$ wavefields from the $\mathbf{u}_{n}, \mathbf{v}_{n}$ by either
(a) sequentially applying the transform matrix $k$ times as in traditional timestepping or
(b) calculating the effects of the transform matrix to the $k$th power and applying it to $\mathbf{u}_{n}$ and $\mathbf{v}_{n}$.

\subsection{Second-order partial differential equations}
Second-order system of equations typically arise from first-order system of equations. 
In this section we extend the equations for second-order systems based on the learnings from the previous section. 
The corresponding second-order system (in time) derived from equation \ref{eq:fun1} is
\begin{eqnarray} \label{eq:fun13}
  \left[
  \begin{array} {c}
  \mathbf{\ddot{u}} \\
  \mathbf{\ddot{v}} 
  \end{array}
  \right]
  =
  \begin{bmatrix}
    0 & A \\
    B & 0 
  \end{bmatrix}
  \left[
  \begin{array} {c}
  \mathbf{\dot{u}} \\
  \mathbf{\dot{v}} 
  \end{array}
  \right]
  =
  \begin{bmatrix}
    0 & A \\
    B & 0 
  \end{bmatrix}^{2}
  \left[
  \begin{array} {c}
  \mathbf{u} \\
  \mathbf{v} 
  \end{array}
  \right] 
  =
  \begin{bmatrix}
    A B & 0 \\
    0 & B A 
  \end{bmatrix}
  \left[
  \begin{array} {c}
  \mathbf{u} \\
  \mathbf{v} 
  \end{array}
  \right] ~,
\end{eqnarray}
recalling that $A$ and $B$ are time invariant. 
Note that the $A$ and $B$ block-matrices are in the main diagonal of the system matrix, while they were in off-diagonal position for the first-order system. Using equation \ref{eq:fun3}, we can propose the finite difference solution of equation \ref{eq:fun13} as follows:
\begin{eqnarray} \label{eq:fun14}
  \left[
  \begin{array} {c}
  \mathbf{u}_{n+1} \\
  \mathbf{v}_{n+1} 
  \end{array}
  \right]
  =
  \begin{bmatrix}
    I & A \\
    B & (I+BA) 
  \end{bmatrix}^2
  \left[
  \begin{array} {c}
  \mathbf{u}_{n-1} \\
  \mathbf{v}_{n-1}
  \end{array}
  \right] ~.
\end{eqnarray}
We show next that equation \ref{eq:fun14} can be transformed to the traditional second-order finite-difference schemes. Specifically we need to show that $\mathbf{v}_{n+1}$ in equation \ref{eq:fun14} can be written only as a function of $\mathbf{v}_{n}$ and $\mathbf{v}_{n-1}$ with no $\mathbf{u}_{n}$ and $\mathbf{u}_{n-1}$ contributions. The bottom line of equation \ref{eq:fun3},
\begin{equation} \label{eq:fun15}
  \mathbf{v}_{n+1} = B \mathbf{u}_n +(I + BA) \, \mathbf{v}_n ~,
\end{equation}
can be brought to the following form by substituting the second line in equation \ref{eq:fun2}
to yield
\begin{equation} \label{eq:fun16}
  \mathbf{v}_{n+1} = 2 \, \mathbf{v}_n - \mathbf{v}_{n-1} + BA \, \mathbf{v}_n ~.
\end{equation}
In conventional FD schemes, this equation would have been used to directly discretize equation \ref{eq:fun13}. Note that the same derivation could have been shown for $\mathbf{u}_{n+1}$ using equations \ref{eq:fun2} and \ref{eq:fun3} as shown in the derivation below
\begin{equation} \label{eq:fun17}
  \mathbf{u}_{n+1} = \mathbf{u}_n + A \mathbf{v}_n = \mathbf{u}_n + A(\mathbf{v}_{n-1} + B \mathbf{u}_n) = \mathbf{u}_n + \mathbf{u}_n - \mathbf{u}_{n-1} + AB \, \mathbf{u}_n ~.
\end{equation}

The second-order discretization scheme contains 3 time steps, $\mathbf{v}_{n-1}$, $\mathbf{v}_n$ and $\mathbf{v}_{n+1}$.
The following system advances the time steps by one step and contains a square matrix
\begin{eqnarray} \label{eq:fun18}
  \left[
  \begin{array} {c}
  \mathbf{v}_{n} \\
  \mathbf{v}_{n+1} 
  \end{array}
  \right]
  =
  \begin{bmatrix}
    0 & I \\
    -I & 2I+BA 
  \end{bmatrix}
  \left[
  \begin{array} {c}
  \mathbf{v}_{n-1} \\
  \mathbf{v}_n 
  \end{array}
  \right]
  =
  \begin{bmatrix}
    0 & I \\
    -I & G 
  \end{bmatrix}
  \left[
  \begin{array} {c}
  \mathbf{v}_{n-1} \\
  \mathbf{v}_n 
  \end{array}
  \right] ~,
\end{eqnarray}
where for simplicity of notation we have designated $G=2I+BA$, where $G=G(k=1)$ is the block matrix
containing a single-step Green's function in each row as seen later.
Squaring the system matrix advances the time steps by two, from $\mathbf{v}_{n-1}$, $\mathbf{v}_n$ steps to $\mathbf{v}_{n+1}$, $\mathbf{v}_{n+2}$ steps
\begin{eqnarray} \label{eq:fun19}
  \left[
  \begin{array} {c}
  \mathbf{v}_{n+1} \\
  \mathbf{v}_{n+2} 
  \end{array}
  \right]
  =
  \begin{bmatrix}
    0 & I \\
    -I & G 
  \end{bmatrix}^2
  \left[
  \begin{array} {c}
  \mathbf{v}_{n-1} \\
  \mathbf{v}_n 
  \end{array}
  \right]
  =
  \begin{bmatrix}
    -I & G \\
    -G & G^2 - I 
  \end{bmatrix}
  \left[
  \begin{array} {c}
  \mathbf{v}_{n-1} \\
  \mathbf{v}_n 
  \end{array}
  \right] ~,
\end{eqnarray}
where the squared system matrix is symplectic (Hairer \cite{hairer}).
We can generalize the system to $k$ arbitrary time steps by bringing the system matrix to the $k$th power: 
\begin{eqnarray} \label{eq:fun20}
  \left[
  \begin{array} {c}
  \mathbf{v}_{n+k-1} \\
  \mathbf{v}_{n+k} 
  \end{array}
  \right]
  =
  \begin{bmatrix}
    0 & I \\
    -I & G 
  \end{bmatrix}^k
  \left[
  \begin{array} {c}
  \mathbf{v}_{n-1} \\
  \mathbf{v}_n 
  \end{array}
  \right] ~,
\end{eqnarray}
Note that the structure of the second-order system matrix corresponds to the structure of the system matrix for the first-order system.
Comparing equation \ref{eq:fun19} with equation \ref{eq:fun9}, the differences are:
(i) in the second-order system the main diagonal has a $-I$ component rather than $I$ as it corresponds to the second to 
last time step, (ii) the off-diagonal terms are skew symmetric.
As in the case of the first-order system, we can calculate the $\mathbf{v}_{n+k-1}$ and $\mathbf{v}_{n+k}$ steps by either 
(a) sequentially applying the transform matrix k times as in traditional timestepping or 
(b) calculating the effects of the transform matrix to the k-th power and applying it to $\mathbf{v}_{n-1}$ and $\mathbf{v}_n$.

Similarly to the first-order systems, these transform matrices are invertible indicating forward and backward time symmetry
  $\begin{bmatrix}
    0 & I \\
    -I & G 
  \end{bmatrix}^{-1}
  =
  \begin{bmatrix}
    G & -I \\
    I & 0 
  \end{bmatrix}$
  ~.

\section{System matrix representations as propagator matrices}
In this section we relate the system matrices to physical PMs and describe the wavefield propagation 
in terms of dot products applications of PMs to current state variables (wavefields).

Denote the resulting first- and second order systems from equations \ref{eq:fun9} and \ref{eq:fun20} as
\begin{eqnarray} \label{eq:gf1}
  \left[
  \begin{array} {c}
  \mathbf{u}_{n+k} \\
  \mathbf{v}_{n+k} 
  \end{array}
  \right]
  =
  \begin{bmatrix}
    I & A \\
    B & I+BA 
  \end{bmatrix}^k
  \left[
  \begin{array} {c}
  \mathbf{u}_n \\
  \mathbf{v}_n 
  \end{array}
  \right] 
  = 
  \begin{bmatrix}
  S_{11}(k) & S_{12}(k) \\
  S_{21}(k) & S_{22}(k) 
  \end{bmatrix}
  \left[
  \begin{array} {c}
  \mathbf{u}_n \\
  \mathbf{v}_n 
  \end{array}
  \right] ~,
\end{eqnarray}
and
\begin{eqnarray} \label{eq:gf2}
  \left[
  \begin{array} {c}
  \mathbf{v}_{n+k-1} \\
  \mathbf{v}_{n+k} 
  \end{array}
  \right]
  =
  \begin{bmatrix}
    0 & I \\
    -I & G 
  \end{bmatrix}^k
  \left[
  \begin{array} {c}
  \mathbf{v}_{n-1} \\
  \mathbf{v}_n 
  \end{array}
  \right] 
  = 
  \begin{bmatrix}
  S_{11}(k) & S_{12}(k) \\
  S_{21}(k) & S_{22}(k) 
  \end{bmatrix}
  \left[
  \begin{array} {c}
  \mathbf{v}_{n-1} \\
  \mathbf{v}_n 
  \end{array}
  \right] ~,
\end{eqnarray}
where in either case $\bm{S}(k)$ denotes the $k^{th}$ superstep propagator matrix with the pre-calculated block matrices
\begin{eqnarray} \label{eq:gf3}
  S(k) =
  \begin{bmatrix}
  S_{11}(k) & S_{12}(k) \\
  S_{21}(k) & S_{22}(k) 
  \end{bmatrix} ~.
\end{eqnarray}
Each block matrix $S_{ij}(k)$ corresponds to the $i^{th}$ output state variable vector and the $j^{th}$ input 
state variable vector. 
The size of block matrix $S_{ij}(k)$ is $m \times q$ where $m$ is the length of the $i^{th}$ output state 
variable vector and $q$ is the length of the $j^{th}$ input state variable vector. 
To better understand $S(k)$ we need to look at the structure of the individual block matrix products such as 
\begin{eqnarray} \label{eq:gf4}
  \mathbf{\hat{u}}_{n+k} = S_{11}(k) \; \mathbf{u}_{n} 
\end{eqnarray} 
where $\mathbf{\hat{u}}_{n+k}$ is the top-left product in equation \ref{eq:gf1}. 
Equation \ref{eq:gf4} is a discretized form of linear integral kernel operators. 
In wave equations these kernel operators correspond to PMs and the functions these 
kernels operate on are the time source functions. 
Therefore, $S_{11}(k)$ is a matrix of PMs, each PM forming one row of the matrix 
and $\mathbf{u}_{n}$ is the time source function.

A sketch of equation \ref{eq:gf4} is shown indicating contributions to the output
\begin{eqnarray} \label{eq:gf5}
  \begin{bmatrix}
  \hat{u}_0 \\
  \hat{u}_1 \\
  \hat{u}_2 \\
  \vdots 
  \end{bmatrix}_{n+k}  =
  \begin{bmatrix*}[l]
  \leftarrow G_0(k)\rightarrow \; ~ \; &  \cdots  \\
  \;\;\;\;\;\;\;\;\;\;   \leftarrow G_1(k) \rightarrow \;  &  \\
  \;\;\;\;\;\;\;\;\;\;\;\;\;\;\;\;\;\;\;\; \leftarrow G_2(k)  \rightarrow &  \\
  \;\;\;\;\;\;\;\; \vdots &   \ddots
  \end{bmatrix*}
  \begin{bmatrix}
  u_0 \\
  u_1 \\
  u_2 \\
  \vdots
  \end{bmatrix}_n ~.
\end{eqnarray}
where $G_i(k)$ denotes non-zero segments of each propagator matrix that corresponds to $k$ timesteps. 
Note that the dimension of the block matrix rows indicates the physical dimension of the discretized earth models, 
that is $i = N_x * N_y * N_z$ for a 3D earth model where $N_x, N_y, N_z$ are the dimensions in 
the 3 earth model dimensions. 
The non-zero elements are clustered around the main diagonal. 
For example, when $k=0$ (no superstep), the matrix turns into a unit matrix and the output vector is 
the same as the input vector. 
As superstepping size $k$ in $S(k)$ grows, the non-zero elements in each row also grow. 

The $G_i(k)$ non-zero segments in each column of the block matrices constitute the propagator matrix  
corresponding the unit input for each element in the input vector calculated by solving the PDE for $k$ time steps. 
This requires knowing the earth model parameters only and therefore can be pre-computed for each column and 
stored for subsequent use. 
The reciprocity rules between input and output applies and the column based propagator matrices also serve 
as propagator matrices for each row (Wapenaar \cite{Wapenaar2022}).

\begin{figure}[ht!]
  \centering
  \begin{subfigure}[b]{0.6\linewidth}
  \includegraphics[width=\linewidth]{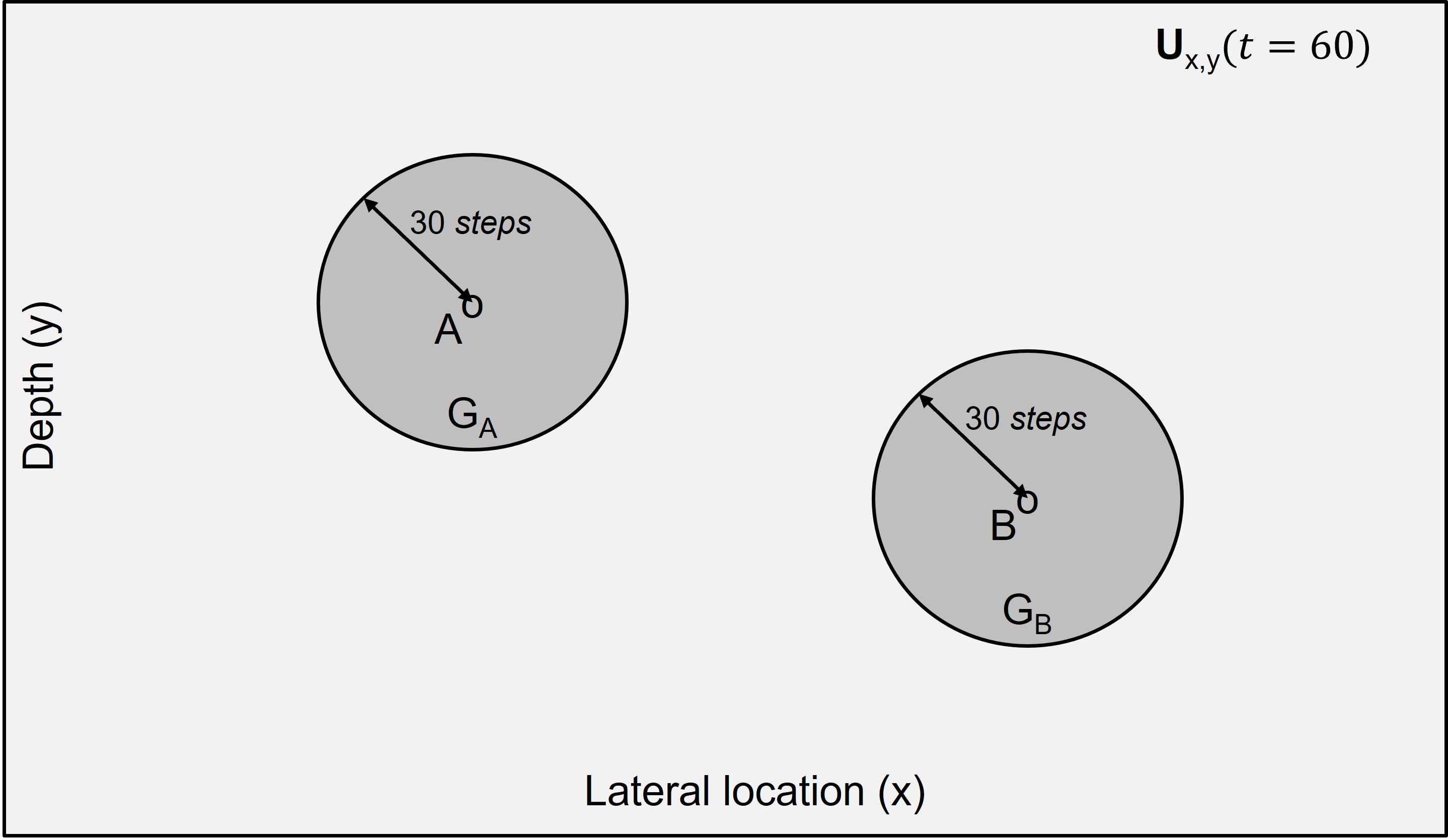}
  \end{subfigure}
  \caption{Sketch illustrating the propagating matrices (PM) in their physical domain. 
  2 PMs $G(A)$ and $G(B)$ shown corresponding to 30 timesteps of 
  propagation from locations $A$ and $B$. 
  The compact support for $G(A)$ and $G(B)$ are shown as gray squares.
  The wavefield $u_{60}$ at time step 60 is indicated in light gray shade and it is computed 
  for the whole physical domain.}
  \label{fig:num2}
\end{figure}

Since the block matrix is sparse, we can represent the matrix-vector multiply in a more compact form, 
keeping only the $G_i(k)$ non-zero segments of the matrix that are pre-computed Green's functions and 
the spatially corresponding subset of $\mathbf{u}_{n}$ denoted as $\mathbf{\tilde{u}}_{n}$, written 
for each row $i$ (that corresponds to a given spatial point $x,y,z$ in the model):
\begin{eqnarray} \label{eq:gf8}
  \hat{u}_i(n+k) = G_i(k) \; \tilde{u}_i(n) ~.
\end{eqnarray}
Equation \ref{eq:gf8} corresponds to the dot product calculation for each vector element $i$. 
A 2D illustration of the scheme is shown in Figure \ref{fig:num2} where 2 PMs $G(A)$ and $G(B)$ 
shown corresponding to 30 timesteps of propagation from locations $A$ and $B$. 
The compact support for $G(A)$ and $G(B)$ are shown as gray squares. 
The wavefield $\mathbf{u}(t=60)$ at time step 60 is indicated in light gray shade and it is computed for 
the whole physical domain.

\begin{figure}[ht!]
  \centering
  \begin{subfigure}[b]{0.6\linewidth}
    \includegraphics[width=\linewidth]{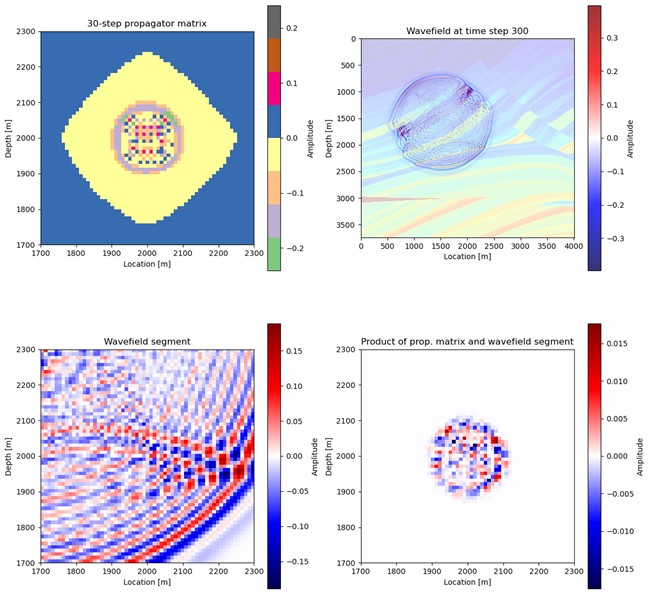}
  \end{subfigure}
  \caption{Illustration of the dot product calculation between a propagating matrix and the corresponding wavefield.
  Top left: PM after 30 time steps. Top right: Wavefield after 300 timesteps 
  in the physical 2D domain. Bottom left: Wavefield segment corresponding to the physical area of the PM. 
  Bottom right: Element-wise multiplication of the PM and the wavefield segment.}
  \label{fig:num3}
\end{figure}

Figure \ref{fig:num3} shows the dot product calculation between the PM (top left) and 
the corresponding wavefield segment (bottom left) resulting in the element-wise products (bottom right). 
Then, these element-wise products are all summed together to generate the resulting scalar of the dot product. 
Note that we used color plots in the top row of Figure \ref{fig:num3} to illustrate the extent of the 
computations: the yellow regions are beyond where the clearly formed wavefront is touched computationally, 
yet it is near-zero value and corresponds to physically acausal area.

The PMs $G_i(k)$ for individual $i$ element are computed independently.
There is overlap in $\tilde{u}_i(n)$ for different $i$ values (locations). 

From this we can describe the computation of of equations \ref{eq:gf1} and \ref{eq:gf2} as
\begin{algorithm}[H]
  \caption{Superstep Wavefield Propagation for Acoustic Wave Equation}
  \label{alg:alg1}
  \begin{algorithmic}[1]
      \Function{SuperstepAcoustic}{\, $k, n_0, N_t, N_{mod}, \mathbf{u}(n_0), \mathbf{v}(n_0), G(k)$, $H(k)$, $K(k)$, $L(k) \,$}
          \For{$n=n_0:k:N_t$\;}  \Comment{iterate in superstep time}
          \For{$i=1:N_{mod}$\;}  \Comment{iterate in model location}
            \State $G_i(k), H_i(k), K_i(k), L_i(k) = \textsc{GetPropMatrix} \left( i \right)$   \Comment{Propagator matrices for the i-th location}
            \State $\tilde{u}_{i}(n)$, $\hat{u}_{i}(n) = \textsc{CutWavefieldSegment} \left( \mathbf{u}(n) \right) $ \Comment{Cut matching wavefield segments}
            \State $\tilde{v}_{i}(n)$, $\hat{v}_{i}(n) = \textsc{CutWavefieldSegment} \left( \mathbf{v}(n) \right) $
            \State $u_i(n+k) = G_i(k) \cdot \tilde{u}_i(n) + H_i(k) \cdot \tilde{v}_i(n)$  \Comment{update $u_i$}
            \State $v_i(n+k) = K_i(k) \cdot \hat{u}_i(n) + L_i(k) \cdot \hat{v}_i(n)$ \Comment{update $v_i$}
          \EndFor
          \EndFor
          \State \textbf{return} $\mathbf{u}, \mathbf{v}$ \Comment{return computed wavefield components}
      \EndFunction
  \end{algorithmic}
\end{algorithm}
where $N_t = T / (\Delta t)$ is the number of time steps for $T$ recording time, 
$\Delta t$ is the temporal sampling rate, 
$k$ is the superstep size,
$n_0$ is the initial time sample level for superstepping, 
$N_{mod} = N_x*N_y*N_z$ is the dimension for 3D-sampled models,   
$G_i(k), H_i(k), K_i(k), L_i(k)$ are propagator matrices at location $i$ corresponding to 
input-output pairings of $u_i \rightarrow u_i, v_i \rightarrow u_i, u_i \rightarrow v_i, v_i \rightarrow v_i$, respectively,
$\tilde{u}_{i}(n), \tilde{v}_{i}(n), \hat{u}_{i}(n), \hat{v}_i(n)$ are the part of the $\mathbf{u}(n)$
and $\mathbf{v}(n)$ state variables (wavefields) that correspond to the compact size of the 
PMs above.
Function $\textsc{GetPropMatrix}$ returns all the PMs for a given model location, and
function $\textsc{CutWavefieldSegment}$ isolates the wavefield segments from the overall wavefield 
that matches the spatial extent of their corresponding PM.
The updated variables $u_i(n+k)$ and $v_i(n+k)$ at time step $n+k$ are the sum of the dot products
of PMs and the corresponding state variables at the $n$-th time step.

We can generalize the number of state variables for more than 2. 
While $\mathbf{u}$ and $\mathbf{v}$ as described in equation \ref{eq:fun1} is valid,
these vectors can be multi-component themselves.
For example, $\mathbf{u}$ can be single-component if it represents a scalar variable such as pressure,
or 3-component vector $\mathbf{u}=(\mathbf{u}_x, \mathbf{u}_y, \mathbf{u}_z)$ if it represents a gradient,
or a 6-component vector $\mathbf{u}=(\mathbf{u}_{xx}, \mathbf{u}_{yy}, \mathbf{u}_{zz}, \mathbf{u}_{xy}, \mathbf{u}_{xz}, \mathbf{u}_{yz})$ 
if it represents the components of more complex differential operators.
The number of components can be different for $\mathbf{u}$ and $\mathbf{v}$ as seen in the examples shown in the Appendices.
The full wavefield $\mathbf{w}$ is a combination of wavefields $\mathbf{u}$ and $\mathbf{v}$ and has $N_c$
components where $N_c$ is the sum of the components in $\mathbf{u}$ and $\mathbf{v}$.
At any location $i$ at any time step $n$ there are $N_c$ state variable values and we can 
represent them by denoting them as $w_{i\alpha}$.
For scalar-component $\mathbf{u}$ and $\mathbf{v}$ this corresponds to $N_c=2$ with
$\mathbf{w}_{0} = \mathbf{u}$ and $\mathbf{w}_{1} = \mathbf{v}$.
Since the state variables are $N_c$-component for both the input and the output, 
the corresponding system matrix is the size of $N_c \times N_c$ block matrices.
Therefore we will need to calculate and use $N_c \times N_c$ propagator matrices for each location.
These PMs at a given location corresponds to the various combinations of
input and output components.
We will denote these PMs as $G_{i\alpha\beta}(k)$ where $i$ is the index of the physical location,
$\alpha$ is the index of the output component, $\beta$ is the index of the input component
and $k$ is the superstep size.
These generalizations can be written as:
\begin{algorithm}[H]
  \caption{Superstep Wavefield Propagation for General Wave Equations}
  \label{alg:alg2}
  \begin{algorithmic}[1]
      \Function{SuperstepGeneral}{\, $k, n_0, N_t, N_{mod}, N_c, \mathbf{w}(n_0), G(k)$}
          \For{$n=n_0:k:N_t$\;}     \Comment{iterate in superstep time}
          \For{$i=1:N_{mod}$\;}     \Comment{iterate in model location}
          \For{$\alpha=1:N_c$ \;}   \Comment{iterate for output wavefield components}
          \For{$\beta=1:N_c$ \;}    \Comment{iterate for input wavefield components}
            \State $G_{i\alpha\beta}(k) = \textsc{GetPropMatrix} \left( i \right)$   \Comment{Propagator matrices for the i-th location}
            \State $\tilde{w}_{i\alpha\beta}(n) = \textsc{CutWavefieldSegment} \left( \mathbf{w}(n) \right) $ \Comment{Cut matching wavefield segments}
          \EndFor  
          \State $w_{i\alpha}(n+k) = \sum_{\beta=1}^{N_c} G_{i\alpha\beta}(k) \cdot \tilde{w}_{i\alpha\beta}(n)$ \Comment{update for $w_{i\alpha}$}          
          \EndFor
          \EndFor
          \EndFor
          \State \textbf{return} $\mathbf{w}$ \Comment{return computed wavefield components}
      \EndFunction
  \end{algorithmic}
\end{algorithm}

We can also generalize Algorithm \ref{alg:alg2} to multiple survey records (whole survey) by defining
tensor $w_{i\alpha j}$ where each column $w_{i\alpha}$ is a single record as above
and where index $j$ indicates the index of the record:
\begin{algorithm}[H]
  \caption{Multi-survey Superstep Wavefield Propagation for General Wave Equation}
  \label{alg:alg3}
  \begin{algorithmic}[1]
      \Function{SuperstepGeneralMultiSurvey}{\, $k, n_0, N_t, N_{mod}, N_c, N_s, \mathbf{w}(n_0), G(k)$ \,}
          \For{$n=n_0:k:N_t$\;}     \Comment{iterate in superstep time}
          \For{$i=1:N_{mod}$\;}     \Comment{iterate in model location}
          \For{$\alpha=1:N_c$ \;}   \Comment{iterate for output wavefield components}
          \For{$j=1:N_{s}$\;}     \Comment{iterate in surveys}
          \For{$\beta=1:N_c$ \;}    \Comment{iterate for input wavefield components}
            \State $G_{i\alpha\beta}(k) = \textsc{GetPropMatrix} \left( i \right)$   \Comment{Propagator matrices for the i-th location}
            \State $\tilde{w}_{i\alpha\beta j}(n) = \textsc{CutWavefieldSegment} \left( \mathbf{w}(n) \right) $ \Comment{Cut matching wavefield segments}
          \EndFor  
          \State $w_{i\alpha j}(n+k) = \sum_{\beta=1}^{N_c} G_{i\alpha\beta}(k) \cdot \tilde{w}_{i\alpha\beta j}(n)$ \Comment{update for $w_{i\alpha j}$}          
          \EndFor
          \EndFor
          \EndFor
          \EndFor
          \State \textbf{return} $\mathbf{w}$ \Comment{return computed wavefield components}
      \EndFunction
  \end{algorithmic}
\end{algorithm}
where $N_s$ indicates the total number of surveys.
Note that there is no dependence on PM $G_{i\alpha\beta}(k)$ from a particular record $j$.
As readily seen above, the fundamental task, the unit of computation is the dot product between a PM
and a state variable wavefield component of the same size.
The other tasks are to get the precomputed PMs 
and to select the appropriate wavefield components that correspond to those PMs.
Thus we can build a computational graph based on these tasks to account for the total
computation used during the simulation of the survey. 
The optimized computational graph determines the execution order of the nested loop elements.

\section{Superstepping computational model}
Superstepping creates two fundamentally different tasks to propagate wavefields:
(a) Precompute - creating PMs for a given model and
(b) Compute - using these PMs to advance the wavefields.
In this section we examine the computational consequences of these very different tasks.

\subsection{Precompute: Propagator matrix generation}
In equation \ref{eq:gf5}, the system matrix consists of rows of PMs (flattened for multidimensional PM) 
and the input and output vectors represent the current and updated wavefields. 
Since the material (earth model) parameters are all contained in the system matrix,  
therefore the PMs for each spatial location represent all the physics applied in wavefield propagation,  
including the physical model encoded in the partial differential equations, 
all the numerical issues of physics such as dispersion, boundary conditions.  
In addition, the distribution of the PMs in space  
and the number of PMs at each spatial location can also be considered information about the physics. 

Since the system matrices in equations \ref{eq:gf1} and \ref{eq:gf2} are fixed, 
the system is linear and time invariant. 
As a consequence, the PMs can be precomputed and applied to all measurement time ranges and all shot records. 
This presents an opportunity to generate high quality PMs where all the desired physics is incorporated. 
The PMs can also be efficiently generated in a non-overlapping parallel way for typical model sizes. 

\begin{figure}[ht!]
  \centering
  \begin{subfigure}[b]{0.6\linewidth}
    \includegraphics[width=\linewidth]{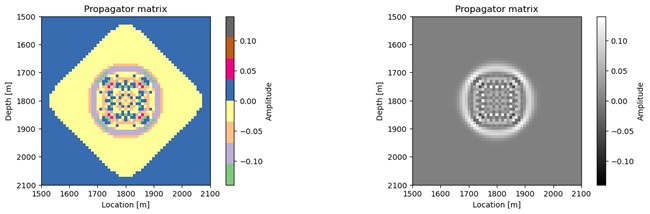}
  \end{subfigure}
  \caption{Comparison between the numerical and physical domains of influence. 
  The same 30-step propagator matrix is plotted with "Accent" color scale in (a) and
  with gray in (b). The yellow diamond in (a) is the computational domain while the formed
  ring of wave in (b) is the physical domain.}
  \label{fig:num4}
\end{figure}

One way to accelerate the computation of PMs is to reduce their size to their physical domain of influence. 
As seen on Figure \ref{fig:num4}a, the computational domain of influence is shown in the yellow diamond 
with the physical wavefront already formed in the middle of the diamond. 
The corner areas in blue are untouched by computation after these time steps. 
We can reduce the size of the PMs by trimming around the wavefront while also preserving numerical precision. 
This can be thought of bringing two different parts of physics into the PMs before discarding physics: 
(a) computing the PM coefficients based on physics and
(b) determining the causal regions based on both math and physics,
further tightening the compact support of PMs.

\begin{figure}[ht!]
  \centering
  \begin{subfigure}[b]{0.6\linewidth}
    \includegraphics[width=\linewidth]{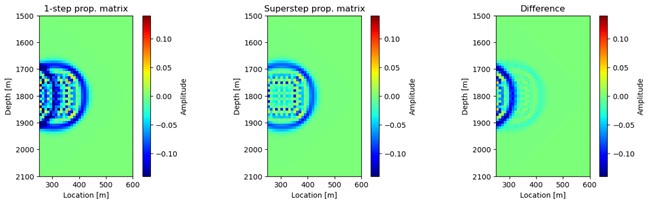}
  \end{subfigure}
  \caption{A propagator matrix near a reflecting boundary (a) without compensation for the boundary layer; 
  (b) compensated for the boundary layer and (c) their difference.}
  \label{fig:num5a}
\end{figure}

Another way to improve the computed PMs is to incorporate the physics of handling 
boundary reflections into them. 
Figure \ref{fig:num5a} shows a PM example with boundary reflections still in it (\ref{fig:num5a}a), 
boundary reflections removed from it (\ref{fig:num5a}b) and their difference (\ref{fig:num5a}c). 
In the Numerical examples section below we demonstrate the effect with and without boundary reflections.

We can reduce dispersion in the PMs by employing high fidelity numerical schemes during their computation. 
Since PMs are shared in different time scales and among all survey records, 
the optimization efforts provide significant value for production runs. 
Although in our examples we ran 2nd order spatial numerical schemes, in production we would replace them 
with higher spatial order schemes. 

PMs represent the filter that moves the incoming wavefields $k$ timesteps forward
essential for computing gradients for full waveform inversion or migration images in reverse time migration. 
These same PMs can also be used to move the incoming wavefield $k$ timesteps backward. 
In practice this corresponds to taking the inverse of matrix in equation \ref{eq:gf3} 
and we illustrate this in Appendix E.

PMs can be further improved by generating representations to reduce grid effects or 
improving the wavefield continuity. 
These representations can be accomplished by either traditional signal processing or machine learning. 
While we do not show examples of these representations in this paper, it is well-understood 
that higher quality PMs generate better wavefields.

PMs can also be used for approximations of the full wave equations, such as rays, dynamic rays 
or other simpler geometrical optics approximations as smooth gradients with second order derivatives.

\subsection{Compute: wavefield propagation}
Wavefield propagation in superstepping is agnostic to physics as physics has already been parameterized and 
incorporated into PMs. 
Therefore one can consider wavefield propagation a computer science task that uses parameterized PMs. 
From this perspective, the implementations need to be general to efficiently map to a broad range of 
compute resources and at the same time allow for efficient representations of PMs and 
their interactions with wavefields.

Once a simulation task has been initiated on a set of compute resources, a computational graph can be created 
that connects PMs, source locations and receiver locations and the simulation time ranges 
with wavefields. 
This graph represents the overall computational effort to connect these components and it can be 
mapped to (distributed) compute resources programmatically using the most efficient computer science methods.

The generality of mapping is facilitated by the simplification of tasks during superstepping:
(a) keeping PMs in memory awaiting wavefields;
(b) select and crop the wavefields to the size of PMs and bring them to the corresponding memory and 
(c) compute the dot product between the PM and the cropped wavefield.
These operations capture the available utilities on a compute node, namely, memory, interconnect and compute resources. 
Maximizing all these resources is thus a computational optimization goal.

Since all of the PMs are available at the wavefield propagation stage, the goal is 
to keep all of them in memory for fast availability. 
For large wavefield propagation tasks, this can be on the order of thousands to tens of thousands of 
compute nodes for a non-optimized representation. 
For example, consider a typical CPU node with over a 100 GB or $10^{11}$ bytes of memory. 
If we perform wavefield propagation on a large $1000^{3}$ gridded model, we would have $10^{9}$ grid locations. 
In each grid location there would be 3 to 45 Green's PMs where, 3 is for acoustic and 45 for full elastic 
models of wave propagation. 
This comes from the physical representation of system matrices as 2x2 block matrices for acoustic and 9x9 block 
matrices for elastic equations. 
Assume the superstep to be $k=30$ and each PM is of the size of up to $30^3$ or 2.7$\times10^4$ bytes. 
Therefore the total size of PMs in a location is in the order of up to $10^6$ bytes. 
To store all these PMs in memory takes $10^{15}$ bytes.  
For this example we would need tens of thousands ($\approx 10^{4}$) nodes to keep all the PMs in memory. 
While this is a very large computational resource for memory, in practice we can optimize the PM 
representation significantly to reduce this to hundreds to thousands of nodes that is the typical size of 
the task using a traditional finite-difference implementation. 
Note that keeping the Green's functions in the memory of many nodes sets up a large distributed computation. 
This is facilitated by that the location (index $i$) is the second outermost loop in Algorithm \ref{alg:alg3}.

The wavefields used for propagation (2 for acoustic and 9 for elastic) are also distributed among the nodes. 
Similarly to the traditional implementations, some halo exchanges of the wavefields happen. 
As each node can host hundreds to hundreds of thousands of locations, these locations can share the 
same wavefields. 

The computational task of the dot product between the PM and the cropped wavefield is less 
intensive compared to the task of managing PMs and wavefields. 
There are several ways to optimize this dot product and to facilitate the above goals. 
First, develop memory-efficient representations of PMs that improves the dot product calculation. 
All PMs can be processed up-front before the dot product computation. 
The coefficients can be flattened and compressed using run-length encoding and kept in memory in a compressed form. 
The dot product calculation can be executed while the PM coefficients are in the 
run-length encoded form. 
This allows more PMs kept in the memory of a single node and it also reduces the number of 
halo exchanges for the wavefields.

The PMs can be also decomposed by bit order and the dot product executed at bit order level 
taking into account the different statistics of the higher bits and the lower bits. 
The multiplication at bit level is just an addition if 1 and no action if 0. 
At the high-end level one can train a machine learning model and use the trained model to infer 
the PM for the dot product computation. 
Second, we can optimize the dot product computation by utilizing that many PMs are kept in 
a single node's memory. 
From the previous example, 45 PMs are kept in memory per location (elastic). 
These can be combined into a single simplified representation where their dot products are computed jointly. 
In the prior example, $10^5$ locations will fit in single node's memory that is, a size of $45^3$ sub-cubes 
can be jointly optimized by forming a reduced representation by truncated SVD or other mathematical techniques.

It is worth mentioning that Algorithm \ref{alg:alg3} prescribes a distributed computing model that is unusual 
from traditional finite-difference modeling perspective. 
It favors parallelization by earth model locations and not by individual shot record locations. 
Note that the earth model locations (loops indexed by $i$) are outer to the shot records (loops indexed by $j$) 
in Algorithm \ref{alg:alg3}.

\section{Synthetic examples}
In this section we illustrate some aspects of superstepping on a subset of the well-known 2D Marmousi model \cite{marmousi}. First, we demonstrate that Green's functions can incorporate the physics of handling boundary reflections without any specific boundary reflection treatment during superstepping. Later we show an example of superstepping on Marmousi model compared to traditional wavefield propagation. 

\begin{figure}[ht]
  \includegraphics[width=0.8\linewidth]{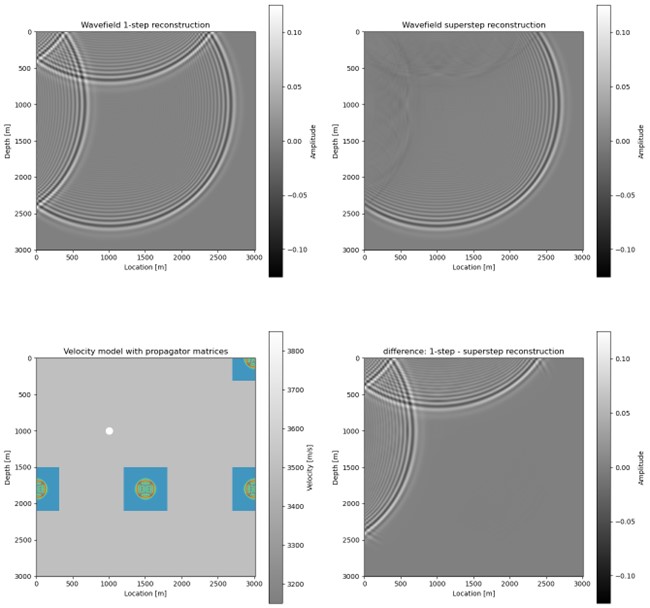}
  \caption{Wavefield after 420 traditional time steps. 
  Top left: Wavefield after 420 steps calculated by traditional method;
  Top right: Wavefield after 420 steps calculated by superstep method. Initial wavefield at 30 steps was advanced by 30 steps 
  at once using propagator matrices;
  Bottom left: Homogeneous velocity model overlaid by propagator matrices at proportional size and the source location (white dot);
  Bottom right: Difference between the 1-step and superstep methods at the time scale of the wave field.
  }
  \label{fig:num5}
\end{figure}

Figure \ref{fig:num5} shows a wavefield propagating in a homogeneous medium with 3500 m/s velocity.
The model size is 301x301 with 10m spatial sampling.
The source is at location (1000,1000), denoted by a white dot in the bottom left figure.
Several PMs are shown over the model indicating the relative size of PMs after 30 regular time steps.
The top left figure shows the wavefield after 420 traditional single-step time stepping with no boundary condition applied.
Strong boundary reflections are observed in the top and left of the model.
The top right figure shows the wavefield after superstepping using the PMs in Figure \ref{fig:num4}
where the boundary conditions were addressed during the PM calculation.
The PMs were calculated for $k=30$ superstep size.
A total of 13 supersteps were applied to the initial wavefield at 30 traditional steps.
No boundary conditions and boundary zones were defined and used during superstepping.
The bottom right figure shows the difference between the wavefield in the top left and right figures.
As seen, the difference corresponds to the boundary reflections.

The superstepping procedure can also be interpreted in terms of Huygens' principle.
All points of the wavefield $\mathbf{u}(n)$ after $n$ time steps act as point sources that scatter waves for $k$ time steps. 
The sum of the scattered waves result in the full wavefield at time step $n + k$. 
Including both $\mathbf{u}(n)$ and $\mathbf{u}(n-1)$ in the superstepping
procedure provides the temporal direction so that the full wavefield propagates forward in time.
Equivalently, the superstepping procedure can be interpreted as a gathering operation.
In this case, for a homogeneous medium with pressure field, 
the PMs $\mathbf{G}(k)$ after $k$ time steps are hollow spheres.
The zero lag correlation between $\mathbf{G}(k)$ and $\mathbf{u}(n)$ (vector dot product) is collected
at the PM source locations to yield wavefield $\mathbf{u}(n+k)$,
the only nonzero contribution from the superstepping procedure 
corresponds to PMs tangent to $\mathbf{u}(n)$. 
Thus, the wavefield is propagated by supersteps of size $k$.
For a virtual boundary separating the domain of interest from an infinite domain, PMs
can be truncated by the boundary either partially or totally if their center falls outside the domain of interest.
As a result, the wavefield portion that falls outside the domain of interest will also be truncated 
without having any effect on the one remaining inside.
Considering the wavefield inside the domain of interest, truncating the PMs by the virtual boundary is
equivalent to propagating the wavefield in the unbounded medium (without the boundary). 
Therefore, PM truncation effectively simulates Sommerfeld
radiating condition, e.g., the wavefield only exists the domain (outward propagation) without reentry.
For more complex media with the possibility of back-scattering from the external region,
the effective internal regions needs to be enlarged until the backscattering can be neglected.

\begin{figure}[ht]
  \includegraphics[width=0.8\linewidth]{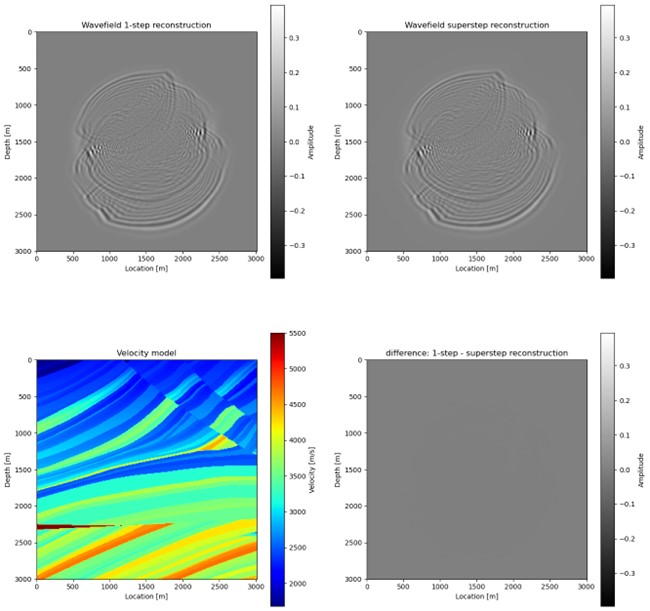}
  \caption{Wavefield after 300 traditional time steps. 
  Top left: Wave field after 300 steps calculated by traditional method;
  Top right: Wavefield after 300 steps calculated by superstep method. Initial wavefield at 30 steps was advanced by 30 steps 
  at once using propagator matrices;
  Bottom left: Velocity model (301x301 subset of Marmousi);
  Bottom right: Difference between the 1-step and superstep methods at the time scale of the wave field.
  }
  \label{fig:num6}
\end{figure}

\begin{figure}[ht]
  \includegraphics[width=0.8\linewidth]{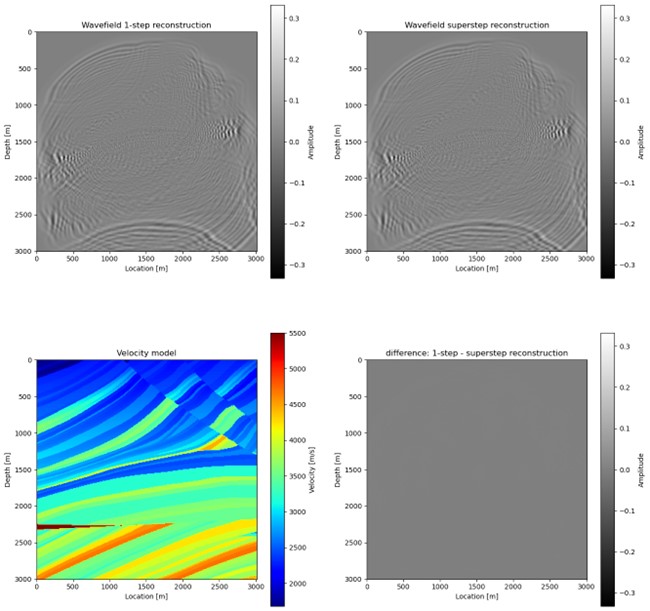}
  \caption{Wavefield after 450 traditional time steps. Sub-figures are same as in Figure \ref{fig:num5}. 
    Please note that the boundary edge reflections are accurately reproduced by the superstep method.}
  \label{fig:num7}
\end{figure}

The superstepping scheme was applied in a similar manner to a 300x300 grid section of the Marmousi model
shown in the bottom left figure in Figure \ref{fig:num6}.
The source location is at the middle of the model at (1500m, 1500m).
The top left figure shows the wavefield after 300 traditional time steps and 
the top right shows the corresponding superstep-propagated wavefield after 9 30-traditional-step supersteps were applied to the initial wavefield.
The bottom left figure shows the difference between the wavefields in the top left and right figures.
As expected, the superstep results closely matched the results achieved using traditional time stepping.
Figure \ref{fig:num7} shows the wavefields after 450 traditional time steps or 14 superstep after the initial 30 traditional time steps.
No boundary layers were used and no boundary reflection compensation included in the PMs either.
As a result, both the traditional and the superstep-generated wavefields contains significant boundary reflections
and the boundary reflections are identical.
This result further confirms that superstepping does not contain additional physics that is not already included in the PMs.

\newpage
\section{Discussion}
We have developed a superstepping scheme for linear or linearized partial differential equations where 
the effect of the repeated application of the system matrix can be precomputed in an efficient manner.
This result has several important consequences.

The superstep formulation is a general scheme that applies to all all discretized PDEs that are linear w.r.t. the solution.
For example, Algorithm \ref{alg:alg3} is applicable to acoustic or elastic wave equation, with or without
density terms. It is also applicable to propagate wavefields for other, more complex discretized equations.
The difference between the various equations manifests in the content of the PMs and the number 
of those PMs for a given location.
In the Appendices we gave examples for first and second-order acoustic schemes and VTI and elastic first-order formulations.

The superstep formulation clearly separates tasks that are associated with the physics and task associated 
with wavefield propagation. 
This was our original goal when we initiated this effort.
The first task can be precomputed and the resulting PMs used in a later wavefield propagation.
We have shown that all physics can be accounted for during the generation of the PMs
and these PMs can be further enhanced.
We have shown that the second, the wavefield propagation task is well-suited for computer science optimization
and compute resource mapping.
Specifically, the three steps in wavefield propagation:
(a) keeping PMs in memory awaiting wavefields; 
(b) selecting and cropping the wavefields to the size of the PMs and bring them to the corresponding memory 
and (c) compute the dot product between the PM and the cropped wavefield, 
are fully utilizing the resources on compute nodes and
allow computational optimization without the need to know the underlying physics.

Propagator matrices are computed for each grid location in the earth models and the locations
are manifested in the outer loops in Algorithm \ref{alg:alg3}.
This results in a distributed computing scheme where the main distribution factor is 
the placement of sufficient number of PMs in a node's memory.
Efficient memory utilization is the driving force for computational optimization and it
underscores that most large-scale wavefield propagation job are bandwidth (memory or IO) limited.

It may be futile to advance wavefields via superstepping at many traditional timesteps at once if 
in many cases we would need to interpolate back to all time step for imaging (or adjoint-state-based gradient computation) purposes.
We have developed imaging schemes using superstepping that avoid these interpolations
and the wavefield propagation and imaging can proceed with supersteps. 

Since the wavefield propagation is heavily influenced by the computer science optimization,
the future success of this method fully depends on strong domain-specific language implementations.

\section{Acknowledgements}
We are thankful for the Chevron Technical Center for the permission to publish this paper.
Ziyi Yin contributed to this work during an internship at Chevron Technical Center.

\nocite{bn2}
\nocite{bw}
\nocite{bw2}

\nobibliography{foo.bib}


\begin{appendices}
\numberwithin{equation}{section}

\section{Matrix exponential derivation of the first order wave equations}
In this section we we show an alternative derivation of the superstepping scheme by using 
a more general matrix exponential formulation of wave propagation.
The velocity-stress equation
\begin{eqnarray} \label{eq:alt1}
  \left[
  \begin{array} {c}
  \mathbf{\dot{u}} \\
  \mathbf{\dot{v}} 
  \end{array}
  \right]
  =
  \begin{bmatrix}
    0 & A \\
    B & 0 
  \end{bmatrix}
  \left[
  \begin{array} {c}
  \mathbf{u} \\
  \mathbf{v}
  \end{array}
  \right] ~,
\end{eqnarray}
can also be written as
\begin{eqnarray} \label{eq:alt2}
  \mathbf{\dot{w}}
  = H \mathbf{w}   ~,
\end{eqnarray}
where $\mathbf{w}= (\mathbf{u}, \mathbf{v})^T$ and the system matrix denoted by $H$.
The formal solution of this matrix equation is
\begin{eqnarray} \label{eq:alt3}
  \mathbf{w}(t) = e^{tH} \mathbf{w}(0)   ~.
\end{eqnarray}
The matrix exponential can be written in power series:
\begin{eqnarray} \label{eq:alt4}
  e^{tH} = \lim\limits_{k \to \infty} \left( I + \frac{tH}{k} \right)^k = 
  \sum_{m=o}^{\infty} \frac{(tH)^m}{m!} 
  = I + \frac{(tH)}{1!} + \frac{(tH)^2}{2!} + \frac{(tH)^3}{3!} + \dots   
\end{eqnarray}
where $I$ is a 2x2 block identity matrix. 
The power series describes numerical integration from $0$ to $t$ for an arbitrary $t$ time value.
Substituting the specific matrix for $H$ and discretizing time as $t=n\Delta t$, we get
\begin{eqnarray} \label{eq:alt5}
  e^{tH} = 
  \begin{bmatrix}
    I & 0 \\
    0 & I 
  \end{bmatrix} + n \Delta t
  \begin{bmatrix}
    0 & A \\
    B & 0 
  \end{bmatrix} + \frac{(n \Delta t)^2}{2!}
  \begin{bmatrix}
    AB & 0 \\
    0 & BA 
  \end{bmatrix} + \frac{(n \Delta t)^3}{3!}
  \begin{bmatrix}
    0 & ABA \\
    BAB & 0 
  \end{bmatrix} + \frac{(n \Delta t)^4}{4!}
  \begin{bmatrix}
    ABAB & 0 \\
    0 & BABA 
  \end{bmatrix} + \dots
\end{eqnarray}
where $n$ is a natural number and $\Delta t$ is the time step that is chosen based on physical considerations.
Next, we limit the power series for a single time step ($n=1$).
The zeroth order approximation ($m=0$) is limiting as in this case no update happens ($e^{tH} \approx I$).
The first order approximation ($m=1$) is also limiting as neither state variable completes a 
round trip from $\mathbf{u}_{0}$ to $\mathbf{u}_{1}$, or $\mathbf{v}_{0}$ to $\mathbf{v}_{1}$,
e.g., $\mathbf{u}_{0}$ does not contribute to $\mathbf{u}_{1}$,
\begin{eqnarray} \label{eq:alt6a}
  e^{\Delta tH} \approx  
  \begin{bmatrix}
    1 & 0 \\
    0 & 1 
  \end{bmatrix} + \Delta t
  \begin{bmatrix}
    0 & A \\
    B & 0 
  \end{bmatrix}
\end{eqnarray}
resulting in the following partitioned explicit Euler finite-difference scheme
\begin{eqnarray} \label{eq:alt6b}
  \mathbf{u}_{n+1} & = & \mathbf{u}_n + A \, \mathbf{v}_{n} ~, \nonumber \\
  \mathbf{v}_{n+1} & = & \mathbf{v}_n + B \, \mathbf{u}_{n}   ~.
\end{eqnarray}
It is the second order approximation ($m=2$) where such a round trip happens first
\begin{eqnarray} \label{eq:alt6}
  e^{\Delta tH} \approx  
  \begin{bmatrix}
    1 & 0 \\
    0 & 1 
  \end{bmatrix} + \Delta t
  \begin{bmatrix}
    0 & A \\
    B & 0 
  \end{bmatrix} + \frac{(\Delta t)^2}{2!}
  \begin{bmatrix}
    AB & 0 \\
    0 & BA 
  \end{bmatrix}  
\end{eqnarray}
but this partitioned implicit Euler method is not symplectic and not used in practice.
Instead, the most common practice is to use symplectic operator splitting methods, 
such as Lie-Trotter, Strang and higher order composition methods (Hairer \cite{hairer}).
If we remove one of the block matrices from the second order term in equation \ref{eq:alt6} 
and compensate for it by multiplying the second oder term by 2, 
we create a partitioned explicit-implicit Euler approximation that is symplectic and
can be split by using the Lie-Trotter splitting scheme
\begin{eqnarray} \label{eq:alt7}
  e^{\Delta tH} \approx  
  \begin{bmatrix}
    I & \Delta t A \\
    \Delta t B & I + (\Delta t)^2 BA 
  \end{bmatrix} =
  \begin{bmatrix}
    I & 0 \\
    \Delta t B & I 
  \end{bmatrix} 
  \begin{bmatrix}
    I & \Delta t A \\
    0 & I 
  \end{bmatrix} =
  e^{\Delta tH_B} \circ e^{\Delta t H_A} ~.
\end{eqnarray}
This is the approximation we used in equation \ref{eq:fun2}.

With $n$ time steps we need $n$ round trips for state variables so for an appropriately chosen $\Delta t$
this corresponds to applying the Lie-Trotter split-step operators chained $n$ times
\begin{eqnarray} \label{eq:alt8}
  e^{n \Delta tH} \approx ( e^{\Delta tH_B} \circ e^{\Delta tH_A} )^n ~.
\end{eqnarray}

\section{First-order system for the acoustic wave equation}
In this section we derive the first-order system matrix for the acoustic  wave equation.
This wave equation is expressed as based on equation \ref{eq:fun1}:
\begin{eqnarray}  \label{eq:appFirstOrder1}
    \left[
    \begin{array} {c}
        \dot{\mathbf{v}} \\
        \dot{p} 
    \end{array}
    \right]
    =
    \begin{bmatrix}
            0 &  -1/\rho \; \nabla \\
            - \kappa \nabla \cdot &  0 
    \end{bmatrix}
    \left[
    \begin{array} {c}
        \mathbf{v} \\
        p
    \end{array}
    \right]  ~,
\end{eqnarray}
where $\mathbf{v}$ is the velocity vector, $p$ is pressure, $\rho$ is density
and $\kappa$ is the adiabatic compression modulus.
Equation \ref{eq:appFirstOrder1} is typically discretized the following way:
\begin{eqnarray}  \label{eq:appFirstOrder2}
    \mathbf{v}_{n+1/2} & = & \mathbf{v}_{n-1/2} - \Delta t/\rho \; \nabla \, p_n  ~, \nonumber \\
    p_{n+1} & = & p_n - \Delta t \; \kappa \nabla \cdot \, \mathbf{v}_{n+1/2}  ~,
\end{eqnarray}
Note that $\mathbf{v}$ and $p$ are on staggered grid relative to each other and the
material properties are sampled on their respective grid.
Expressing the $k$ time step shifted components $\mathbf{v}_{n-1/2+k}$ and $p_{n+k}$ 
from $\mathbf{v}_{n-1/2}$ and $p_n$ yields
\begin{eqnarray}  \label{eq:appFirstOrder3}
    \left[
    \begin{array} {c}
        \mathbf{v}_{n-1/2+k} \\
        {p}_{n+k}
    \end{array}
    \right]
    =
    \begin{bmatrix}
            I &  - \Delta t/\rho \; \nabla \\
            - \Delta t \; \kappa \nabla \cdot &  I+ (\Delta t)^2 \;\kappa \nabla \cdot ( 1/\rho \; \nabla )
    \end{bmatrix}^k
    \left[
    \begin{array} {c}
        \mathbf{v}_{n-1/2} \\
        p_n
    \end{array}
    \right]  ~,
\end{eqnarray}
where the system matrix is now the superstep propagator matrix in equation \ref{eq:gf3}:
\begin{eqnarray}  \label{eq:appFirstOrder4}
    \left[
    \begin{array} {c}
        \mathbf{v}_{n-1/2+k} \\
        {p}_{n+k}
    \end{array}
    \right]
    = 
    \begin{bmatrix}
        S_{11}(k) & S_{12}(k)  \\
        S_{21}(k) & S_{22}(k) 
    \end{bmatrix}
    \left[
    \begin{array} {c}
        \mathbf{v}_{n-1/2} \\
        p_n
    \end{array}
    \right]  ~.
\end{eqnarray}
The last equation is formally equivalent to equation \ref{eq:gf1}.

\section{First-order system for the elastic wave equation}
In this section we derive the first-order system for the elastic wave equation.
The elastic wave equation is expressed as based on \ref{eq:fun1}:
\begin{eqnarray}  \label{eq:appFirstOrderElastic1}
    \left[
    \begin{array} {c}
        \dot{v}_x \\
        \dot{v}_y \\
        \dot{v}_z \\
        \dot{\sigma}_{xx} \\
        \dot{\sigma}_{yy} \\
        \dot{\sigma}_{zz} \\
        \dot{\sigma}_{yz} \\
        \dot{\sigma}_{xz} \\
        \dot{\sigma}_{xy} 
    \end{array}
    \right]
    =
    \begin{bmatrix}
            0 &  1/\rho \; \nabla_{3 \times 6} \\
            C \; \nabla_{6 \times 3} &  0 
    \end{bmatrix}
    \left[
    \begin{array} {c}
        {v}_x \\
        {v}_y \\
        {v}_z \\
        {\sigma}_{xx} \\
        {\sigma}_{yy} \\
        {\sigma}_{zz} \\
        {\sigma}_{yz} \\
        {\sigma}_{xz} \\
        {\sigma}_{xy} 
    \end{array}
    \right]  ~,
\end{eqnarray}
where $\nabla_{3 \times 6}$ and $\nabla_{6 \times 3}$ are the matrix differential operators
\begin{eqnarray} \label{eq:appFirstOrderElastic2}
  \nabla_{3 \times 6} =
  \begin{bmatrix}
  \partial_x & 0 & 0 & 0 & \partial_z & \partial_y \\
  0 & \partial_y & 0 & \partial_z & 0 & \partial_x \\
  0 & 0 & \partial_z & \partial_y & \partial_x & 0
  \end{bmatrix} ~; \;\;\;
  \nabla_{6 \times 3} =
  \begin{bmatrix}
  \partial_x & 0 & 0 \\
  0 & \partial_y & 0 \\
  0 & 0 & \partial_z \\
  0 & \partial_z & \partial_y \\
  \partial_z & 0 & \partial_x \\
  \partial_y & \partial_x & 0
  \end{bmatrix} ~;
\end{eqnarray}
and $\mathbf{v} = [v_x, v_y, v_z]^T$ is the particle velocity vector, 
$[\sigma_{xx}, \sigma_{yy}, \sigma_{zz}, \sigma_{yz}, \sigma_{xz}, \sigma_{xy}]^T$
are the six independent components of the stress tensor $\bm{\sigma}$
and bold upper case $\bf{C}$ will denote the $6 \times 6$ matrix of stiffnesses 
in the Voigt notation \cite{chapman}.
After discretization of equation \ref{eq:appFirstOrderElastic1} and taking $k$ traditional time steps
(that is, raising the system matrix to the $k$-th power), the state vectors at the $n+k$ step are 
related to the state vectors at the $n$-th steps as
\begin{eqnarray}  \label{eq:appFirstOrderElastic3}
    \left[
    \begin{array} {c}
        {v}_x \\
        {v}_y \\
        {v}_z \\
        {\sigma}_{xx} \\
        {\sigma}_{yy} \\
        {\sigma}_{zz} \\
        {\sigma}_{yz} \\
        {\sigma}_{xz} \\
        {\sigma}_{xy} 
    \end{array}
    \right]_{n+k}
    =
    \begin{bmatrix}
            I &  \Delta t /\rho \; \nabla_{3 \times 6} \\
            \Delta t \; C \; \nabla_{6 \times 3}  & I+ (\Delta t)^2 C \; \nabla_{6 \times 3} (1/\rho \; \nabla_{3 \times 6})
    \end{bmatrix}^k
    \left[
    \begin{array} {c}
        {v}_x \\
        {v}_y \\
        {v}_z \\
        {\sigma}_{xx} \\
        {\sigma}_{yy} \\
        {\sigma}_{zz} \\
        {\sigma}_{yz} \\
        {\sigma}_{xz} \\
        {\sigma}_{xy} 
    \end{array}
    \right]_n  ~,
\end{eqnarray}
where the time step value of the state vectors is denoted at the bottom tight corner of their square bracket.
Here the system matrix is now the superstep propagator matrix in equation \ref{eq:gf3}
\begin{eqnarray}  \label{eq:appFirstOrderElastic4}
    \left[
    \begin{array} {c}
        {v}_x \\
        {v}_y \\
        {v}_z \\
        {\sigma}_{xx} \\
        {\sigma}_{yy} \\
        {\sigma}_{zz} \\
        {\sigma}_{yz} \\
        {\sigma}_{xz} \\
        {\sigma}_{xy} 
    \end{array}
    \right]_{n+k}
    =
    \begin{bmatrix}
        S_{11}(k) & S_{12}(k)  \\
        S_{21}(k) & S_{22}(k) 
    \end{bmatrix}
    \left[
    \begin{array} {c}
        {v}_x \\
        {v}_y \\
        {v}_z \\
        {\sigma}_{xx} \\
        {\sigma}_{yy} \\
        {\sigma}_{zz} \\
        {\sigma}_{yz} \\
        {\sigma}_{xz} \\
        {\sigma}_{xy} 
    \end{array}
    \right]_n  ~,
\end{eqnarray}
and last equation is formally equivalent to equation \ref{eq:gf1}.

\section{A 5 $\times 5$ first-order system for the VTI acoustic wave equation}
In this section we derive the 5 $\times 5$ first-order system for the VTI acoustic wave equation.
The 5 $\times 5$ first-order system is based on \ref{eq:fun1}:
\begin{eqnarray} \label{eq:appFirstOrderVTI1}
  \left[
  \begin{array} {c}
  \dot{v}_x \\
  \dot{v}_y \\
  \dot{v}_z \\
  \dot{\sigma}_{H} \\
  \dot{\sigma}_{M}
  \end{array}
  \right]
  =
  \begin{bmatrix}
    0 & A \\
    B & 0 
  \end{bmatrix}
  \left[
  \begin{array} {c}
  {v}_x \\
  {v}_y \\
  {v}_z \\
  {\sigma}_{H} \\
  {\sigma}_{M} 
  \end{array}
  \right] ~,
\end{eqnarray}
where
\begin{eqnarray} \label{eq:appFirstOrderVTI2}
  A = 
  \begin{bmatrix}
  (1/\rho) \; \partial_x & 0 \\
  (1/\rho) \; \partial_y & 0 \\
  \frac{\partial_z}{\rho (1+2\eta) \sqrt{1+2\delta}} & \frac{\sqrt{2\eta}\partial_z}{\rho (1+2\eta) \sqrt{1+2\delta}}
  \end{bmatrix} ~, \;\;\; \nonumber \\ \\
  B = 
  \begin{bmatrix}
  \rho V^2_{Pz} (1+2\varepsilon) \partial_x & \rho V^2_{Pz} (1+2\varepsilon) \partial_y & \rho V^2_{Pz} \sqrt{1+2\delta} \; \partial_z \\
  0 & 0 & \rho V^2_{Pz} \sqrt{2\eta} \sqrt{1+2\delta} \; \partial_z 
  \end{bmatrix} ~, \nonumber
\end{eqnarray}
$\mathbf{v} = [v_x, v_y, v_z]^T$ is the particle velocity vector, 
$\bm{\sigma} =[\sigma_{H}$, $\sigma_{M}]^T$ are the horizontal and mixed horizontal and vertical  components of the stress tensor $\bm{\sigma}$,
$\rho$ is density, $\varepsilon$, $\delta$ and $\eta$ are the anisotropy parameters.
For details of definitions and derivations of equation \ref{eq:appFirstOrderVTI2} see Bube et al. \cite{bn2}.
Equation \ref{eq:appFirstOrderVTI1} is typically discretized the following way:
\begin{eqnarray}  \label{eq:appFirstOrderVTI3}
    \mathbf{v}_{n+1/2} & = & \mathbf{v}_{n-1/2} + \Delta t \; A \; \bm{\sigma}_n  ~, \nonumber \\
    \bm{\sigma}_{n+1} & = & \bm{\sigma}_n + \Delta t \; B \; \mathbf{v}_{n+1/2}  ~,
\end{eqnarray}
Note that $\mathbf{v}$ and $\bm{\sigma}$ are on staggered grid relative to each other and the
material properties are sampled on their respective grid.
Expressing the $k$ time step shifted components $\mathbf{v}_{n-1/2+k}$ and $\bm{\sigma}_{n+k}$ 
from $\mathbf{v}_{n-1/2}$ and $\bm{\sigma}_n$ yields
\begin{eqnarray}  \label{eq:appFirstOrderVTI4}
    \left[
    \begin{array} {c}
        {v}_x \\
        {v}_y \\
        {v}_z \\
        {\sigma}_{H} \\
        {\sigma}_{M}  
    \end{array}
    \right]_{n+k}
    =
    \begin{bmatrix}
        I & \Delta t \; A  \\
        \Delta t \; B & I + (\Delta t)^2 BA
    \end{bmatrix}^k
    \left[
    \begin{array} {c}
        {v}_x \\
        {v}_y \\
        {v}_z \\
        {\sigma}_{H} \\
        {\sigma}_{M}  
    \end{array}
    \right]_n  ~,
\end{eqnarray}
where for simplicity we ignored the staggered grid notation when denoting the state vector time steps.
Here the system matrix is also the superstep propagator matrix in equation \ref{eq:gf3}
\begin{eqnarray}  \label{eq:appFirstOrderVTI5}
    \left[
    \begin{array} {c}
        {v}_x \\
        {v}_y \\
        {v}_z \\
        {\sigma}_{H} \\
        {\sigma}_{M}  
    \end{array}
    \right]_{n+k}
    =
    \begin{bmatrix}
        S_{11}(k) & S_{12}(k)  \\
        S_{21}(k) & S_{22}(k) 
    \end{bmatrix}
    \left[
    \begin{array} {c}
        {v}_x \\
        {v}_y \\
        {v}_z \\
        {\sigma}_{H} \\
        {\sigma}_{M}  
    \end{array}
    \right]_n  ~,
\end{eqnarray}
and last equation is formally equivalent to equation \ref{eq:gf1}.

\section{Second-order system for the acoustic wave equation}
The second-order system for the acoustic wave equation is expressed as based on \ref{eq:fun16}:
\begin{equation} \label{eq:appSecondOrder1}
  \ddot{p} = V^2 \Delta p 
\end{equation}
where $p$ is pressure and $V$ is P-wave velocity.
Equation \ref{eq:appSecondOrder1} is typically discretized the following way:
\begin{equation} \label{eq:appSecondOrder2}
  p_{n+1} = 2 \, p_n - p_{n-1} + \Delta t \; V^2 \Delta p_n = G p_n - p_{n-1}
\end{equation}
where $G=G_1=2+\Delta t \; V^2 \Delta$ is a single step operator. 
In general, let $G_k$ to be the propagator matrix for advancing $k$ time levels, 
taking into account that we need to store two adjacent time levels to be able to advance in time for second-order systems.
$G_k$ is defined by 
\begin{equation} \label{eq:appSecondOrder3}
  G_k (p_0) = p_k ~,
\end{equation}
where $p_k$ is the solution of equation \ref{eq:appSecondOrder2} at $n=k$ with initial conditions $p_{-1} = 0$ and $p_0$.

The $k^{th}$ time step based on \ref{eq:fun20}: 
\begin{eqnarray} \label{eq:appSecondOrder4}
  \left[
  \begin{array} {c}
  p_{n+k-1} \\
  p_{n+k} 
  \end{array}
  \right]
  =
  \begin{bmatrix}
    0 & I \\
    -I & G 
  \end{bmatrix}^k
  \left[
  \begin{array} {c}
  p_{n-1} \\
  p_n 
  \end{array}
  \right] ~,
\end{eqnarray}
or
\begin{eqnarray} \label{eq:appSecondOrder5}
  \left[
  \begin{array} {c}
  p_{n+k-1} \\
  p_{n+k} 
  \end{array}
  \right]
  =
  \begin{bmatrix}
  S_{11}(k) & S_{12}(k) \\
  S_{21}(k) & S_{22}(k)
  \end{bmatrix}
  \left[
  \begin{array} {c}
  p_{n-1} \\
  p_n 
  \end{array}
  \right] ~,
\end{eqnarray}
where the last equation is formally equivalent to equation \ref{eq:gf2}.

Next, we show derive properties of $\bm{S}(k)$ and $G_k$ that will indicate reverse time propagation using
these rotated block matrices.
Block matrix multiplication from equation \eqref{eq:appSecondOrder4} gives
\begin{eqnarray} \label{eq:aappSecondOrderB1}
  S(2) = S(1)^2 =
  \begin{bmatrix}
    -I & G \\
    -G & G^2 -I
  \end{bmatrix} ~,
\end{eqnarray} 
\begin{eqnarray} \label{eq:appSecondOrderB2}
  S(3) = S(1)^3 =
  \begin{bmatrix}
    -G & G^2 - I \\
    -(G^2-I) & G^3 -2 G
  \end{bmatrix} ~.
\end{eqnarray} 
For $k\geq 1$, since $G$ is independent of $k$, we see from the definition of $G_k$ that the lower
right block of $S(k)$ is $G_k$: $p_{n+k}$ would be $G_k(p_n)$ if $p_{n-1}$ were zero. Similarly, the upper right
block of $S(k)$ is $G_{k-1}$: $p_{n+k-1}$ would be $G_{k-1}(p_n)$ if $p_{n-1}$ were zero. 
For $k\geq 2$, consider the solution of equation \eqref{eq:appSecondOrder2} with a given $p_{n-1}$ and zero $p_n$.
By equation \eqref{eq:appSecondOrder2}, this solution has zero $p_n$ and $p_{n+1} = -p_{n-1}$; advancing this solution $k-1$
time steps, $p_{n+k}$ would be $G_{k-1}(p_{n+1}) = -G_{k-1}(p_{n-1})$ and $p_{n+k-1}$ would be $G_{k-2}(p_{n+1}) =
-G_{k-2}(p_{n-1})$.
Thus the lower left block of $S(k)$ is $-G_{k-1}$, the upper left block of $S(k)$ is $-G_{k-2}$, and 
for an arbitrary $k$ time step we get
\begin{eqnarray} \label{eq:appSecondOrderB3}
  S(k) = 
  \begin{bmatrix}
    -G_{k-2} & G_{k-1} \\
    -G_{k-1} & G_k 
  \end{bmatrix} ~.
\end{eqnarray}
Now we derive the recurrence relation for $G_k$.
Consider time step $S(k+1)$
\begin{eqnarray} \label{eq:appSecondOrderB4}
  S(k+1) = 
  \begin{bmatrix}
  -G_{k-1} & G_{k} \\
  -G_{k} & G_{k+1} 
  \end{bmatrix} 
  = S(1) S(k) =
  \begin{bmatrix}
  0 & I \\
  -I & G_{1} 
  \end{bmatrix}
  \begin{bmatrix}
  -G_{k-2} & G_{k-1} \\
  -G_{k-1} & G_k 
  \end{bmatrix}
  =
  \begin{bmatrix}
  -G_{k-1} & G_{k} \\
  G_{k-2}-G_1 G_{k-1} & G_1 G_k - G_{k-1} 
  \end{bmatrix}
\end{eqnarray}
we conclude that for $k\geq 1$
\begin{equation} \label{eq:appSecondOrderB5}
  G_{k+1} = G_1 G_k - G_{k-1} ~.
\end{equation}
Another form of equation \ref{eq:appSecondOrderB5} is
\begin{equation} \label{eq:appSecondOrderB6}
  G^2_{k+1} - G_{k-2} G_k = I ~,
\end{equation}
that can be shown directly by induction using the recurrence relation \ref{eq:appSecondOrderB5} twice.

\begin{figure}[h]
  \includegraphics[width=0.95\linewidth]{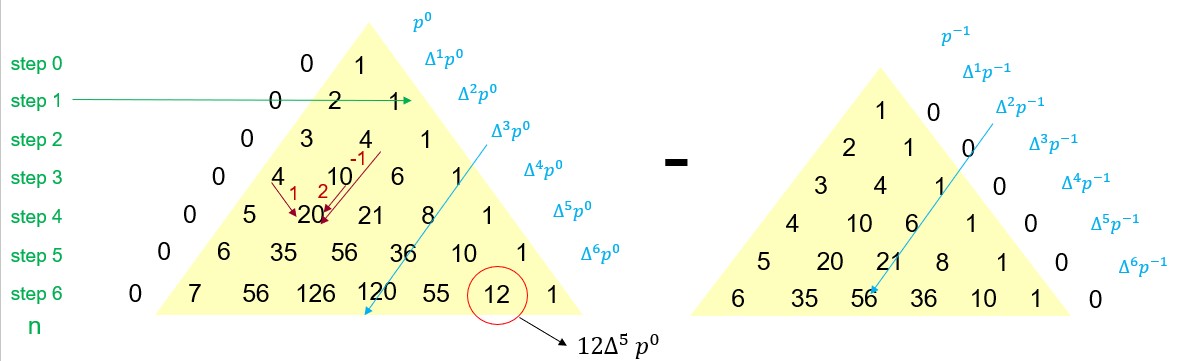}
  \caption{Wave equation triangle}
  \label{fig:AppC2}
\end{figure}

Starting with $G_0 = I$ and $G_1=2+V^2 \Delta$, we get for $G_k$
\begin{eqnarray} \label{eq:appSecondOrderB7}
  \begin{aligned}
  G_2 = G_1^2 - &I ~, \\
  G_3 = G_1^3 - &2 G_1 ~, \\
  G_4 = G_1^4 - &3 G_1^2 + I ~,
  \end{aligned}
\end{eqnarray}
etc. We see that each $G_k$ is a polynomial $q_k(G_1)$ in $G_1$, where $q_k$ is a monic polynomial
of degree $k$: $q_0(x) = 1$, $q_1(x) = x$, $q_2(x) = x^2-1$, $q_3(x) = x^3-2x$, and in general,
$q_{k+1}(x) = xq_k(x)-q_{k-1}(x)$ for $k \geq 1$.
In particular, all of the $G_k$’s commute with each other: for $k \geq 0$ and $n \geq 0$,
\begin{equation} \label{eq:appSecondOrderB8}
  G_k G_n = G_n G_k ~.
\end{equation}
This monic polynomial can be illustrated in a triangle similar to Pascal's Triangle (Figure \ref{fig:AppC2}).

Now we show that the inverse of the superstep propagator matrix is the reverse-time superstep propagator.
Consider equation 
\begin{eqnarray} \label{eq:appSecondOrderB9}
  S(1)^{-1} =
  \begin{bmatrix}
    G_1 & -I \\
    I & 0
  \end{bmatrix} ~.
\end{eqnarray}
Notice that $S(1)^{-1}$ is the same as $S(1)$ with the block columns switched and the block rows switched or 180 degrees rotated. 
This observation can be written in matrix form. Let $R$ be the 2 x 2 block matrix
\begin{eqnarray} \label{eq:appSecondOrderB10}
  R =
  \begin{bmatrix}
    0 & I \\
    I & 0
  \end{bmatrix} ~;
\end{eqnarray}
$R$ is a symmetric permutation matrix, and $R^{-1} = R^T = R$. Either from the observation
above about switching block columns and block rows, or by direct computation, we have
\begin{equation} \label{eq:appSecondOrderB11}
  S(1)^{-1} = R S(1) R ~.
\end{equation}
From this, we can compute $S(k)^{-1}$ for $k \geq 2$:
\begin{eqnarray} \label{eq:appSecondOrderB12}
  S(k)^{-1} = (S(1)^k)^{-1} = (S(1)^{-1})^k = (R S(1) R ) ^k = (R S(1) R ) (R S(1) R ) \cdots (R S(1) R ) 
  = (R S(1)^k R ) \nonumber \\ = (R S(k) R ) =
  \begin{bmatrix}
    0 & I \\
    I & 0
  \end{bmatrix}
  \begin{bmatrix}
    -G_{k-2} & G_{k-1} \\
    -G_{k-1} & G_{k} 
  \end{bmatrix}
  \begin{bmatrix}
    0 & I \\
    I & 0
  \end{bmatrix}
  =
  \begin{bmatrix}
    G_{k} & -G_{k-1} \\
    G_{k-1} & -G_{k-2} 
  \end{bmatrix}
\end{eqnarray}
indicating reverse timestep propagation.

\end{appendices}

\end{document}